\documentstyle[11pt,moriond,epsfig]{article}

\def\Journal#1#2#3#4{{#1} {\bf #2}, #3 (#4)}

\def\NPB{{\em Nucl. Phys.} B}
\def\NPA{{\em Nucl. Phys.} A}
\def\PLB{{\em Phys. Lett.}  B}
\def\PRL{\em Phys. Rev. Lett.}
\def\PRD{{\em Phys. Rev.} D}

\def\vepp{\varepsilon'}
\def\vep{\varepsilon}

\def\be{\begin{equation}}
\def\ee{\end{equation}}
\def\ba{\begin{eqnarray}}
\def\ea{\end{eqnarray}}

\def\cL{{\cal L}}
\def\cQ{{\cal Q}}
\def\cA{{\cal A}}

\begin{document}
\vspace*{4cm}
\title{FINAL STATE INTERACTIONS IN $\vepp / \vep$\,{}
\footnote{Contribution to the
XXXVth Rencontres de Moriond on Electroweak Interactions and Unified Theories,
March 11--18, 2000.}
}
\author{ E. PALLANTE }

\address{Facultat de F\'{\i}sica, Universitat de Barcelona, Diagonal
647,\\ E-08028 Barcelona, Spain  }

\maketitle\abstracts{
Strong final state interaction (FSI) effects can play a central role in the 
Standard Model prediction of weak $K\to 2\pi$ matrix elements. Here, I discuss 
how FSI's affect the direct CP violation parameter $\vepp / \vep$ by solving 
the Omn\`es problem for the necessary  $K\to 2\pi$ amplitudes. 
The main results of this analysis have been reported in a previous paper 
\cite{SHORT} and are further discussed in Refs. \cite{FUTURE_PP,FUTURE_PPS}.
The resulting Standard Model prediction  
 $\mbox{Re} ( \vepp / \vep ) = (15\pm 5)\times 10^{-4}$ is in good agreement
with the present experimental world average.}

\section{Introduction}

In a recent paper \cite{SHORT} it was pointed out that strong final state 
interaction (FSI) effects, when properly taken into account, produce 
a large enhancement (about a factor of two) in the Standard Model prediction 
of the direct CP violation parameter $\varepsilon'/\varepsilon$.
FSI effects also play an important role in the observed
$\Delta I =1/2$ rule \cite{DELTA12,KA91}.

At centre--of--mass energies around the kaon
mass, the strong S--wave $\pi$--$\pi$ scattering generates a large
phase-shift difference
$\left(\delta^0_0 - \delta^2_0\right)(m_K^2)= 45^\circ\pm 6^\circ$
between the $I=0$ and $I=2$ partial waves \cite{GM}. 
This effect is taken into account by factoring out those phases
in the usual decomposition of $K\to 2\pi$ amplitudes with definite isospin 
$I=0$ and $I=2$:
\be\label{eq:AIdef} 
{\cal A}_I \,\equiv\, A\left[ K\to (\pi\pi )_I\right]
\,\equiv\, A_I \; e^{i\delta^I_0} \; .
\ee
The presence of such a large phase--shift difference also signals 
a large dispersive FSI effect in the moduli of the isospin amplitudes, since 
their imaginary and real parts are related by analyticity and unitarity. 
The size of the FSI effect can be already roughly 
estimated at one loop in Chiral Perturbation Theory (ChPT), where the 
rescattering of the two pions in the final state produces an enhancement 
of about $40\%$ in the $A_0$ amplitude \cite{KA91,BPP,JHEP}.
However, the fact that the one loop calculation still underestimates 
the observed $\delta_0^0$ phase shift indicates that a further enhancement 
should be produced by higher orders.

It is then clear that FSI contributions  should be included and
their effect resummed in the description of $K\to 2\pi$ decays.
In addition, the rescattering of the two pions in the final state is a purely 
long--distance mechanism, which is fully decoupled from the production 
mechanism of the two pions in the decay process. As a consequence, 
FSI effects do not introduce any dependence on short--distance 
quantities (i.e. they do not carry any dependence on the 
factorization scale in the OPE description of $K\to 2\pi$ amplitudes).

Despite the importance of FSI in $K\to 2\pi$ decays has been known for more 
than a decade, most of the {\it Standard Model predictions}
of $\varepsilon'/\varepsilon$ \cite{Buras,Lattice} did not include those
contributions and failed to  reproduce the experimental measurements 
\cite{exp}. Lattice determinations of 
$K\to 2\pi$ amplitudes are still missing FSI effects, since they are obtained 
in a two--steps procedure where first the simpler $K\to \pi$ matrix element
is measured on the lattice and second, the physical $K\to 2\pi$ matrix elements
are obtained by using a lowest--order ChPT relation  
between $K\to \pi$ and $K\to 2\pi$. Neither the first nor the second step 
include FSI effects. 

On the other hand, approaches based on effective low--energy 
models \cite{Trieste}
or the $1/N_c$ expansion \cite{Dortmund}
do include some one-loop corrections and find larger values for the
$A_0$ amplitude. However, the drawback in these cases could be 
the possible model dependence of the matching procedure with 
short--distance. 

The approach that has been recently proposed \cite{SHORT,FUTURE_PP,FUTURE_PPS}
 is the Omn\`es approach, which permits a 
resummation of the universal FSI effects to all orders in ChPT.
In Section~\ref{OMNES} the general Omn\`es problem is formulated, while in 
Section~\ref{SCALAR} the Omn\`es problem is solved for the pion scalar form 
factor, a well known quantity in ChPT; this analysis clarifies the relevant 
properties of the Omn\`es solution. Finally, in Section~\ref{KAON} we 
consider the Omn\`es solution for the $K\to 2\pi$ amplitudes
 which enter the prediction of $\varepsilon'/\varepsilon$. 
The final result of an exact matching procedure with 
short--distance \cite{FUTURE_PPS}, with the inclusion of FSI 
effects \cite{FUTURE_PP} gives 
\be\label{eq:epsth}
\left.\mbox{Re}\left(\varepsilon'/\varepsilon\right)\right|_{SM} = 
(15\pm 5)\times 10^{-4} \, ,
\ee
which is in good agreement with the present experimental world average
\cite{Ceccuchi}
\be\label{eq:exp}
\left.\mbox{Re}\left(\varepsilon'/\varepsilon\right)\right|_{exp} = 
(19.3\pm 2.4)\times 10^{-4} \, .
\ee

\section{The Omn\`es problem}
\label{OMNES}

Let us consider a generic amplitude (or form factor) $A^I_J(s)$,
with two pions in the final state
which have total angular momentum and isospin given by $J$ and $I$,
respectively, and invariant mass 
$s\equiv q^2\equiv \left(p_1 + p_2\right)^2$.
The amplitude is analytic everywhere except for a cut on the real positive 
$s$ axis $L = [4 m_\pi^2, \infty )$. 

Below the first inelastic threshold, only the $2\pi$ intermediate state 
contributes to the absorptive part of the amplitude and Watson's theorem 
\cite{Watson} implies that the phase of the amplitude is equal to the  
phase of the $\pi\pi$ partial--wave
scattering amplitude, so that
\ba\label{eq:WASIM_EL}
\mbox{Im}\, A^I_J \, =\, (\mbox{Im}\, A^I_J )_{2\pi} &=& 
 e^{-i\delta^I_J} \,\sin{\delta^I_J}\; A^I_J  \, =\, 
e^{i\delta^I_J} \,\sin{\delta^I_J}\; A^{I\ast}_J  
\nonumber\\ &=& 
\sin{\delta^I_J}\; \vert A^I_J\vert \, =\, \tan{\delta^I_J}\; 
\mbox{Re}\, A^I_J\, .
\ea
Cauchy's theorem implies instead that $A^I_J(s)$ can be written
as a dispersive integral along the physical cut:
\be\label{eq:DISP}
A^I_J(s) = {1\over \pi}\int_{L}~ d z \; 
{{\mbox{Im}}\, A^I_J(s)\over z-s-i\epsilon} 
\, +\, {\mbox{subtractions}}\, .
\ee
Inserting Eq.~\ref{eq:WASIM_EL} in the dispersion relation 
\ref{eq:DISP}, one obtains an integral equation for $A^I_J(s)$ of 
the Omn\`es type, 
which has the well--known Omn\`es solution \cite{ALLOMNES}
(for $n$ subtractions with subtraction point $s_0$ outside the physical cut):
\be\label{eq:NSUB} 
A^I_J(s) \, = \, Q^I_{J,n}(s,s_0)~\exp{\left\{I^I_{J,n}(s,s_0)\right\}}
\, ,
\ee
where
\be\label{eq:In}
I^I_{J,n}(s,s_0) \,\equiv\, {(s-s_0)^{n} \over \pi}
\int^\infty_{4m^2_\pi}~ 
{dz\over (z-s_0)^{n}}  \;
{\delta^I_J(z) \over z-s-i\epsilon} \, 
\ee
and 
\be\label{eq:NSUB_Q} 
\log{\left\{Q^I_{J,n}(s,s_0)\right\}} \,\equiv\, \sum_{k=0}^{n-1}\, 
{ (s-s_0)^{k} \over k!}\; {d^k \over d s^k}   
\left.\log{\left\{A^I_J(s)\right\}} \right|_{s=s_0} \, ,
\qquad (n\geq 1)\, ,
\ee
with $Q^I_{J,0}(s,s_0)\equiv 1$.
The dispersive integral $I^I_{J,n}(s,s_0)$ 
is uniquely determined up to a polynomial ambiguity
(that does not produce any imaginary part of the amplitude),
which depends on the number of subtractions and the subtraction point.
The simple iterative relation for the real part of $I^I_{J,n}(s,s_0)$
\be\label{eq:ITER}
\mbox{Re}\,  I^I_{J,n}(s,s_0) \, =\, 
\mbox{Re}\,  I^I_{J,n-1} (s,s_0) \, - \,
(s-s_0)^{n-1}\lim_{s\to s_0} {\mbox{Re}\, I^I_{J,n-1}(s,s_0)
\over (s-s_0)^{n-1}}
\, ,
\ee
shows that only a polynomial part of  $I^I_{J,n}(s,s_0)$ does depend on the 
subtraction point $s_0$ and the number of subtractions $n$, while 
the non--polynomial part of  $I^I_{J,n}(s,s_0)$,
the one containing the infrared chiral logarithms, is universal (i.e. 
$s_0$ and $n$ independent). 
Thus, the Omn\`es solution predicts the chiral logarithmic corrections
in a universal way and 
provides their exponentiation to all orders in the chiral expansion.
The polynomial ambiguity of $I^I_{J,n}(s,s_0)$ and the subtraction 
function  $Q^I_{J,n}(s,s_0)$ can be fixed, at a given order in the 
chiral expansion, by matching the Omn\`es formula (\ref{eq:NSUB}) 
with the ChPT prediction of $A^I_J(s)$. It remains a polynomial ambiguity at 
higher orders.
Notice that in the presence of a zero of the amplitude the Omn\`es solution 
can be found for the factorized amplitude  $\overline{A^I_J}(s)$, such that 
 $A^I_J(s) = (s-\zeta)^p \,\overline{A^I_J}(s)$, where $\zeta$ is a zero of 
order $p$.

\section{The scalar pion form factor}
\label{SCALAR}

The scalar pion form factor  is known up to two loops in ChPT \cite{BGT:98}
and it is a useful quantity to understand the details of the Omn\`es 
solution in the case $I=0$ and $J=0$. It is defined by the matrix element 
of the $SU(2)$ quark scalar density
$\langle\pi^i(p^\prime )\vert \bar{u}u+\bar{d}d\vert\pi^k(p)\rangle 
\,\equiv\, \delta^{ik}\, F_S^{\pi}(t)$. 
At low momentum transfer, ChPT provides a systematic expansion 
of $F_S^{\pi}(t)$ in powers of $t\equiv (p^\prime - p)^2$ 
and the light quark masses \cite{GL:84,GL:85}:
\be
F_S^{\pi}(t) = F_S^{\pi}(0)\left\{ 1+ g(t) +O(p^4)\right\}\, ,
\label{eq:SFF_2}
\ee
where the $O(p^2)$ correction $g(t)$ contains contributions 
from one-loop diagrams and tree--level terms of the $O(p^4)$ ChPT 
lagrangian. It is given by \cite{GL:84,GL:85} ($f\approx f_\pi = 92.4$ MeV):
\ba
g(t) &=& {t\over f^2}\left\{ \left (1-{M_\pi^2\over 2t}\right ) 
\bar{J}_{\pi\pi}(t) + {1\over 4}\,\bar{J}_{KK}(t) + {M_\pi^2\over 18 t}
\,\bar{J}_{\eta\eta}(t)  +4(L_5^r+2L_4^r)(\mu ) \right .\nonumber\\
&&\left.\quad\mbox{} + {5\over 4(4\pi )^2}
\left (\ln{\mu^2\over M_\pi^2} -1\right ) -  {1\over 4(4\pi )^2}
\ln{M_K^2\over M_\pi^2} \right\}\, .
\ea
The counterterms $L_4^r,\, L_5^r$ from the $O(p^4)$ lagrangian are kept
in the numerical analysis at the standard reference value $\mu = M_\rho$, 
where \cite{EC:95} $[L_5^r +2L_4^r](M_\rho) = (0.8\pm 1.1)\times 10^{-3}$. 
The functions $\bar{J}_{\pi\pi}(t)$, $\bar{J}_{KK}(t)$ and 
$\bar{J}_{\eta\eta}(t)$ \cite{GL:84,GL:85} 
are ultraviolet finite and, together with the logarithms, they are produced 
by the one-loop exchange 
of $\pi\pi$, $K\bar K$ and $\eta\eta$ intermediate states. Notice that 
at values of $t$ such that $t\ll 4 M_P^2$
\be
\bar{J}_{PP}(t) = {1\over (4\pi )^2}\left [ {1\over 6} {t\over M_P^2} 
+ {1\over 60} {t^2\over M_P^4} + \ldots \right ]\, ,
\ee 
implying that, below the $P\bar P$ threshold,
$\bar{J}_{PP}(t)$ has a very smooth behaviour and it is strongly suppressed.
In the case of the pion scalar form factor this means that at values of 
$t\ll 4 M_K^2\, , 4 M_\eta^2$ the 
one-loop functions $\bar{J}_{KK}(t)$ and $\bar{J}_{\eta\eta}(t)$
only give small analytic corrections, which are
numerically negligible with respect to the rest of the one--loop corrections.

Let us consider now the Omn\`es solution for $F_S^\pi (t)$
and fix the subtraction polynomial by performing a matching 
with the one-loop ChPT result. This implies at the subtraction point $t_0$
\be
F_S^{\pi}(t)\, =\, F_S^{\pi}(t_0)\cdot \Omega_0 (t,t_0)
\,\approx\,
F_S^{\pi}(0)\,\left\{ 1+ g(t_0)\right\}\cdot \Omega_0 (t,t_0) \, .
\label{eq:OMNES_SFF}
\ee
Thus, knowing the form factor at some low--energy subtraction point
$t_0$, where the momentum expansion can be trusted,
the Omn\`es factor $\Omega_0 (t,t_0)$ provides an evolution of
the result to higher values of $t$, through the exponentiation
of infrared effects related to FSI.
The once--subtracted solution reads:
\be
\Omega_0^{(1)} (t,t_0)
\, =\, \exp\left\{ {t-t_0\over \pi}\int_{4M_\pi^2}^{\bar{z}}\, 
{dz\over z-t_0}\, {\delta_0^0(z)\over z-t-i\epsilon} \right\} 
\,\equiv\, 
\Re_0^{(1)} (t,t_0) \; e^{i\delta^0_0(t)}\, ,
\label{eq:OMNES_F}
\ee
where the integral has been split into its real and 
imaginary part, making explicit that the phase of the Omn\`es factor
is just the original phase-shift $\delta^0_0(t)$, while $\Re_0^{(1)} (t,t_0)$
is its corresponding dispersive factor.
The integral has been cut at the 
upper edge $\bar{z}$, which represents the first inelastic threshold
($\bar{z} = 1$ GeV${}^2$).
In Table~\ref{table:SFF} the resulting value of
$|F_S^{\pi}(M_K^2)/F_S^{\pi}(0)|$ for different subtraction points
$t_0 = 0,\, M_\pi^2,\, 2 M_\pi^2, \, 3 M_\pi^2, \, 4M_\pi^2$ and $M_K^2$
and at $t=M_K^2$ is shown together with the dispersive part of 
the once--subtracted Omn\`es integral $\Re_0^{(1)}(t,t_0)$
and the one loop function $g(t_0)$.
Notice that the Omn\`es factor 
is not defined for $t_0 =M_K^2$, because it lies above the threshold 
of the non--analyticity cut; however,
by its definition in Eq.~(\ref{eq:OMNES_SFF}),
$\Omega (M_K^2,t_0=M_K^2) = 1$ holds.
\begin{table}[htb]
\protect\caption{
The one loop function $g(t_0)$, 
the Omn\`es factor $\Re_0^{(1)}(t,t_0)$ and the modulus of 
$F_S^{\pi}(t)/F_S^{\pi}(t_0)$
are shown at $t=M_K^2$ for different values of the
subtraction point $t_0\in [0,M_K^2]$.
At a given $t_0$, the first value of $|F_S^{\pi}(t)/F_S^{\pi}(t_0)|$
(within brackets) is obtained with
the $O(p^2)$ ChPT prediction for $\delta_0^0$, while the 
experimental phase-shift data has been used in the second one.
The integrals have been cut at $\bar{z} = 1$ GeV${}^2$.} 
\label{table:SFF}
\begin{center}
\begin{tabular}{|c|c|c|c||c|}
\hline
$t_0$  & $g(t_0)$ & 
\multicolumn{2}{c|}{$\Re_0^{(1)} (M_K^2,t_0)$} & 
$|F_S^{\pi}(M_K^2)/F_S^{\pi}(0)|$ \\
 \cline{3-5} & & $O(p^2)$ & exp.  &
$n=1$ \\
\hline
 0           &  0    & 1.23 & 1.45 & 
  (1.23) \ \ 1.45  \\
$M_\pi^2$    & 0.042 & 1.21 & 1.40 &  
  (1.26) \ \ 1.46 \\
$2 M_\pi^2$  & 0.091 & 1.17 & 1.34 & 
  (1.28) \ \ 1.46  \\
$3 M_\pi^2$  & 0.15  & 1.12 & 1.26 & 
  (1.29) \ \ 1.45  \\
$4 M_\pi^2$  & 0.26  & 1.03 & 1.11 &  
  (1.30) \ \ 1.40 \\
$M_K^2$  & $0.54 - 0.46\, i$ & $\equiv 1$ & $\equiv 1$ & 
     1.61  \\
\hline
\end{tabular}
\end{center}
\end{table}
The ChPT calculation of $F_S^\pi(t_0)$ is obviously better
at lower values of $t_0$ where the one-loop correction $g(t_0)$
is smaller. At $t_0 = 4 M_\pi^2$ a sizable 26\% effect is
already found, while at $t_0 = M_K^2$ (the point chosen in the discussion of
the  Omn\`es factor in a recent paper \cite{MARTI})
the correction is so large than one should worry about higher--order 
contributions.
The Omn\`es exponential allows to predict $F_S^\pi(M_K^2)$ in a
much more reliable way, through the evolution of safer results
at lower $t_0$ values.
\begin{table}[htb]
\protect\caption{Dependence of $\Re_0^{(1)} (M_K^2,t_0)$ 
on the upper edge $\bar{z}$ (in GeV${}^2$ units)
of the dispersive integral,
for various choices of $t_0$. 
The fit to the experimental data for $\delta_0^0$
has been used}
\label{table:SFF_Z}
\begin{center}
\begin{tabular}{|c||c|c|c|}
\hline
& \multicolumn{3}{c||}{$\Re_0^{(1)} (M_K^2,t_0)$} 
\\
\cline{2-4} &
$\bar{z}=1$  & $\bar{z}=2$ & $\bar{z}=3$  
\\
\hline
$t_0=0$          & 1.45 & 1.58 & 1.62 
\\
$t_0=M_\pi^2$    & 1.40 & 1.51 & 1.55 
\\
$t_0= 2 M_\pi^2$ & 1.34 & 1.44 & 1.47 
\\
$t_0= 3 M_\pi^2$ & 1.26 & 1.35 & 1.38 
\\
$t_0= 4 M_\pi^2$ & 1.11 & 1.86 & 1.21 
\\
\hline
\end{tabular}
\end{center}
\end{table}
In Table~\ref{table:SFF_Z} the dependence of the Omn\`es integral on 
the upper edge $\bar{z}$ is shown. From those results one can conclude that 
cutting the integral at the first inelastic threshold is actually 
producing an underestimate of the complete FSI effect. 
The sensitivity of the Omn\`es integral to the high--energy region is
reduced by considering a twice--subtracted  
Omn\`es solution \cite{FUTURE_PP}. 

\begin{figure}[htb]
\centerline{\mbox{\epsfxsize=10cm\epsffile{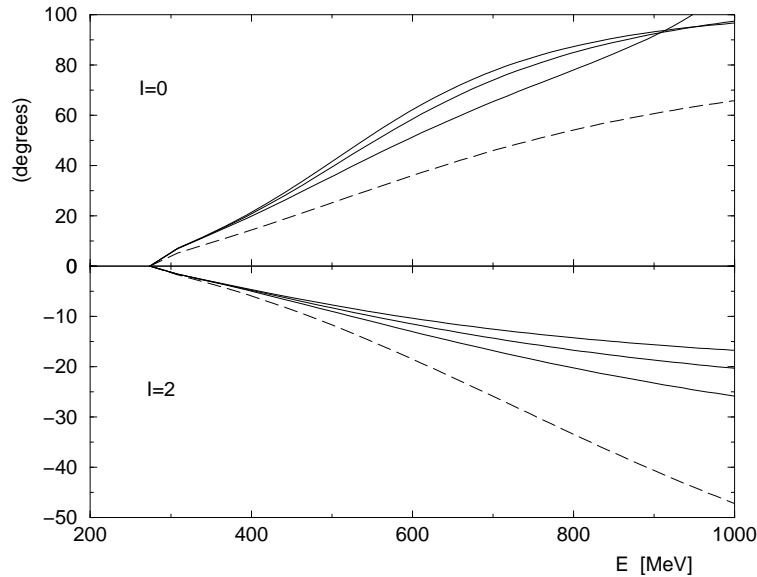}}}
\caption{
Phase shifts $\delta_0^{0;2}(s)$ with $I=0$ and 2, according to a 
fit \protect\cite{SCHENK} of experimental data
and used in the numerical analyses \protect\cite{SHORT,FUTURE_PP}.
Solid lines enclose the range covered by the experimental data, while
dashed lines show the unitarized lowest--order ChPT prediction.}
\label{fig:PHASESHIFT}
\end{figure}

Since chiral corrections are smaller at lower
values of the subtraction point, values of the Omn\`es factor 
at low  $t_0$ should be trusted because less sensitive to higher order 
corrections. Based also on the twice--subtracted result \cite{FUTURE_PP}
one can conclude that the true value of
$|F_S^{\pi}(M_K^2)/F_S^{\pi}(0)|$ is in the range 1.5 to 1.6.
Taking the experimental phase-shift uncertainties into account (see 
Figure~(\ref{fig:PHASESHIFT}) for the details), the result is:
\be
|F_S^{\pi}(M_K^2)/F_S^{\pi}(0)| \, = \, 1.5 \pm 0.1 \, .
\ee
%

\section{Final State Interactions in $\varepsilon' /\varepsilon$}
\label{KAON}

The analysis of the Omn\`es solution for the scalar pion form factor has 
clarified two main points. 
First, the  Omn\`es solution given by the product of the 
polynomial amplitude  times the Omn\`es factor is independent of 
 the subtraction point as it has to be. Second, the  Omn\`es exponential
allows to predict the physical quantity in a much more reliable way, through 
the evolution of safer results at lower values of the subtraction point.

In analogy to the scalar pion form factor, one can derive the Omn\`es 
solution for any $K\to 2\pi$
amplitude, both CP--conserving and CP--violating (the 
presence of a CP--violating weak phase does not modify the 
Omn\`es derivation, being such a phase of short--distance origin 
\cite{FUTURE_PP}).
At lowest order in the chiral expansion, 
the most general effective bosonic Lagrangian, with the same 
$SU(3)_L\otimes SU(3)_R$ transformation properties as the short--distance 
Lagrangian, contains three terms: 
\ba\label{eq:lg8_g27} 
\cL_2^{\Delta S=1} &=& -{G_F \over \sqrt{2}}  V_{ud}^{\phantom{*}} V_{us}^* 
\Bigg\{ g_8  \;\langle\lambda L_{\mu} L^{\mu}\rangle  
 + 
g_{27} \,\left( L_{\mu 23} L^\mu_{11} + {2\over 3} L_{\mu 21} L^\mu_{13} 
\right)  
\Biggr. \nonumber\\ &&\qquad\qquad\quad\mbox{}   + 2 e^2 f^6 g_{EM} \;  
\langle\lambda U^\dagger \cQ U\rangle \Bigg\} + 
\mbox{\rm h.c.} \, .  
\ea 
The flavour--matrix operator $L_{\mu}=-i f^2 U^\dagger D_\mu U$  
represents the octet of $V-A$ currents at lowest order in derivatives,
where $U = exp{i\sqrt{2}\phi/f}$ is the exponential representation of the 
light pseudoscalar meson field with $\phi$ the flavour octet matrix.
$\cQ= {\rm diag}(\frac{2}{3},-\frac{1}{3},-\frac{1}{3})$ is the quark 
charge matrix, 
$\lambda\equiv (\lambda^6 - i \lambda^7)/2$ projects onto the 
$\bar s\to \bar d$ transition [$\lambda_{ij} = \delta_{i3}\delta_{j2}$]
and $\langle A\rangle $ denotes the flavour trace of $A$.

At generic values of the squared 
centre--of--mass energy $s=(p_{\pi 1}+p_{\pi 2})^2$, 
the $I=0,2$ amplitudes generated by the lowest--order lagrangian in Eq.
\ref{eq:lg8_g27} are given by
\ba
A_0 &=& -{G_F\over \sqrt{2}} V_{ud}V^\ast_{us}\,\sqrt{2} f\,
\left\{ \left(g_8+{1\over 9}\, g_{27}\right) (s-M_\pi^2) 
-{4\over 3} f^2 e^2 g_{EM}\right\}  ,
\nonumber\\
A_2&=&  -{G_F\over \sqrt{2}} V_{ud}V^\ast_{us}\, {2\over 9} f
\, \left\{5\, g_{27}\, (s-M_\pi^2)\, - 6 f^2 e^2 g_{EM}\right\} .
\label{TREE}
\ea
In the absence of $e^2 g_{EM}$ corrections,
the tree--level isospin amplitudes have a zero at $s=M_\pi^2$,
because the on-shell amplitudes should vanish in the SU(3) limit
\cite{SU3}. This is not the case for the matrix elements 
of the electroweak penguin operator $Q_8$, whose lowest--order
realization is given by the term proportional to  $e^2 g_{EM}$.
Those matrix elements are constant (of order $e^2 p^0 = p^2$ in the chiral 
power counting) at the lowest order in ChPT. 

For all the cases one can write a once--subtracted Omn\`es solution 
for the on--shell physical amplitudes \cite{SHORT,FUTURE_PP} in the form
\be
\cA_I(M_K^2)  = a_I(M_K^2,s_0)\; \Omega_I^{(1)}(M_K^2,s_0)\, .
\ee
The amplitude $a_I(M_K^2,s_0)$ depends on which weak effective operator 
is mediating the corresponding $K\to 2\pi$ amplitude, while 
the once--subtracted Omn\`es factor  $\Omega_I^{(1)}(M_K^2,s_0)$ 
is universal, 
i.e. the same for the pion scalar form factor $F_S^\pi(s)$ and 
the amplitude $\cA_0(s)$ (this is in principle no more true for the 
twice--subtracted factor that depends on the derivative of the 
function itself). 
Moreover, taking a low subtraction point where higher--order
corrections are very small \cite{FUTURE_PP},
 one can just multiply the tree--level
formulae in Eq. \ref{TREE} with the experimentally determined Omn\`es
exponential. The results are \cite{FUTURE_PP}\footnote{These values 
update earlier estimates of 
$\Re_2^{(1)}(M_K^2,M_\pi^2)$ \cite{ME:91,NS:91}.
The correction factors $\Re_I^{(1)}$ 
were also considered in an estimate
of $\varepsilon'/\varepsilon$ within the 
$SU(2)\otimes SU(2)\otimes U(1)$ model of CP violation
\protect\cite{FRERE}.}
\be\label{eq:R0}
\Re_0(M_K^2,0) = 1.5 \pm 0.1 ~~~~~~~\Re_2(M_K^2,0) = 0.92 \pm 0.06\, ,
\ee
where the subtraction point $s_0=0$ has been chosen.
How the proper inclusion of FSI effects drastically modify the 
Standard Model prediction of 
$\varepsilon^\prime /\varepsilon$ has been already explained \cite{SHORT}.
 
To obtain a complete Standard Model prediction for 
$\varepsilon^\prime /\varepsilon$ an exact matching procedure has been
 proposed \cite{FUTURE_PPS}.
It is  inspired 
by the large--$N_c$ expansion, but only at scales below the charm 
quark mass $\mu \leq m_c$ (where the logarithms that enter the Wilson 
coefficients are small). FSI effects, which are next--to--leading 
in the $1/N_c$ expansion but numerically relevant, are taken into 
account through the multiplicative factors $\Re_I(M_K^2,0)$ while 
avoiding any double counting. 
The general formula for $\varepsilon'/\varepsilon$ can be written as
follows
\be
\mbox{Re}{\varepsilon^\prime\over\varepsilon} = {\omega\over\sqrt{2}
\varepsilon}
{1\over \mbox{Re}A_0}\left ( {1\over\omega} \mbox{Im} A_2 - 
\mbox{Im} A_0\right )\, ,
\label{EPS}
\ee
where $\omega = \mbox{Re}A_2/ \mbox{Re}A_0=1/22.2$ is the inverse of 
the CP--conserving $\Delta I=1/2$ ratio. The CP--conserving quantities
 $\omega$ and $\mbox{Re}A_0$ are taken
from the experiment, while the CP--violating terms $\mbox{Im} A_0$ and 
$\mbox{Im} A_2$ are calculated theoretically. The indirect CP violation 
parameter $\varepsilon$ can be either taken from the experiment or 
evaluated theoretically  \cite{FUTURE_PPS}.
Each contribution to $\mbox{Im} A_0$ and $\mbox{Im} A_2$ 
will now contain the FSI effect 
through the corresponding dispersive factor $\Re_I(M_K^2,0)$ with $I=0$ or 2 
which was not taken into account 
in the previous Standard Model predictions of 
$\varepsilon^\prime /\varepsilon$.
The result \cite{FUTURE_PPS} 
\be
\left.\mbox{Re}\left(\varepsilon'/\varepsilon\right)\right|_{SM} = 
(15\pm 5)\times 10^{-4} \, ,
\ee
has an uncertainty which is dominated by the large--$N_c$ based 
matching procedure. The prediction is within one sigma 
respect to the present experimental world average
\cite{Ceccuchi}
\be
\left.\mbox{Re}\left(\varepsilon'/\varepsilon\right)\right|_{exp} = 
(19.3\pm 2.4)\times 10^{-4} \, ,
\ee
and provides an enhancement of about a factor of two respect to 
the previous short--distance based Standard Model predictions of 
$\varepsilon^\prime /\varepsilon$ \cite{Buras,Lattice}.

\section*{Acknowledgments}
I warmly thank my collaborators Antonio Pich and Ignazio Scimemi
and the organizers of the conference.
 This work has been supported  by the Ministerio de Educaci\'on  y Cultura 
(Spain) and in part by the European Union TMR Network EURODAPHNE 
(Contract No. ERBFMX-CT98-0169).

\end{document}